\documentclass[hidelinks]{scrartcl}
\usepackage{natbib}
\usepackage{float}
\usepackage{amsmath}
\usepackage{enumerate}

\restylefloat{figure}
\usepackage{graphicx}
\usepackage[font=scriptsize,labelfont=bf]{caption}
\usepackage{bookmark}
\usepackage{nameref}
\usepackage{bm}
\usepackage{tikz}
\usetikzlibrary{decorations.pathmorphing}

\makeatletter
\renewcommand*{\@fnsymbol}[1]{\ifcase#1\or\dag \else\@arabic{\numexpr#1-1\relax}\fi}
\makeatother

\providecommand{\keywords}[1]{\noindent\textbf{\textit{Keywords:}} #1}

\begin{document}

	\setlength\intextsep{3pt}
	\newcommand{\smallheading}[1]{\par{\emph{\textbf{#1}}}}
	\title{Synthetic Cognitive Development:}
	\subtitle{\LARGE where intelligence comes from}
	\author{\large David Weinbaum (Weaver) (\texttt{space9weaver@gmail.com})\\
\large Viktoras Veitas (\texttt{vveitas@gmail.com})\footnote{Both authors contributed equally to this work.}\\
\large The Global Brain Institute, VUB}
	\date{\large December 9, 2014}

\maketitle
\pagestyle{plain}

\begin{abstract}
The human cognitive system is a remarkable exemplar of a general intelligent system whose competence is not confined to a specific problem domain. Evidently, general cognitive competences are a product of a prolonged and complex process of cognitive development. Therefore, the process of cognitive development is a primary key to understanding the emergence of intelligent behavior. This paper develops the theoretical foundations for a model that generalizes the process of cognitive development. The model aims to provide a realistic scheme for the synthesis of scalable cognitive systems with an open-ended range of capabilities. Major concepts and theories of human cognitive development are introduced and briefly explored, focusing on the enactive approach to cognition and the concept of sense-making. The initial scheme of human cognitive development is then generalized by introducing the philosophy of individuation and the abstract mechanism of transduction. The theory of individuation provides the ground for the necessary paradigmatic shift from cognitive systems as given products to cognitive development as a formative process of self-organization. Next, the conceptual model is specified as a scalable scheme of networks of agents. The mechanisms of individuation are formulated in context-independent information theoretical terms. Finally, the paper discusses two concrete aspects of the generative model -- mechanisms of transduction and value modulating systems. These are topics of further research towards an implementable architecture.

\end{abstract}

\section*{Introduction}
A primary goal of artificial general intelligence (AGI) research is the synthesis of a machine capable of performing any intellectual task a human being is capable of and eventually going beyond that.  While artificial intelligence research which is problem specific and context specific (e.g. understanding speech and text, visual pattern recognition, robotic motion, various optimization problems, etc.) has lately made quite a few impressive breakthroughs, artificial general intelligence research that aims to distill the principles of intelligence independently of a specific problem domain or a predefined context is still at its preliminary stages. 

The goal of this paper is to present our approach to artificial general intelligence. We frame intelligence in the operation of cognitive agents. The intelligence of a cognitive agent is actualized as a set of competences enabling the agent to effectively respond to problematic situations presented to it by the environment in the course of their interactions. Normally, observing a cognitive system in its operation, we are able to identify a specific problem domain (e.g. motion in a 3D environment), extract the behaviors by which an agent addresses the problem, and create a model that represents these behaviors. The model, if successful, will capture the problem-specific intelligent mechanisms involved and will allow us to apply these mechanisms to similar problems. In such an approach, various organisms and primarily human agents are the exemplars and primary research subjects of intelligent behavior. 

In our investigation, we realized that this approach meets its limits when we aim to understand how intelligent behaviors arise in the first place when an agent meets a problematic situation it has not encountered before and therefore does not possess the \textit{a priori} knowledge of how to address it. Intelligent agents do not appear ready made. Human agents, as one remarkable example, undergo a prolonged and complex process of cognitive development till they become highly competent cognitive systems. If established cognitive competences and their entailed behavioral patterns are the actual manifestations of specific intelligence, it follows quite intuitively that the process of cognitive development by which such cognitive competences arise, is the fulcrum of general intelligence. 

We argue, therefore, that the only effective way to understand general intelligence is by building a model of cognitive development. We describe an abstract genetic\footnote{Genetic in the sense of genesis.} process that can be applied to any general cognitive systems such as biological organisms, human agents, swarms, social systems, social institutions, robots, machine intelligences and more.     

Here, we present a descriptive model that we aim to further develop in the future into a full generative one. The descriptive model provides the principles and conceptual framework of synthetic cognitive development. The goal of the generative model will be to apply our approach to the actual synthesis of systems capable of demonstrating cognitive development, i.e. operating general intelligence in an environment which is not \textit{a priori} framed within a specific problem domain. 

In the first section we specify our approach to cognition as the process of sense-making and identify major characteristics of cognitive development in human agents. In the second section we introduce Simondon's theory of individuation as the philosophical framework for our model. Using the theory of individuation, we depart from the specifics of human cognitive development and re-frame it in a general systemic context. The third section is a description of the conceptual model in information-theoretic terms. The final section outlines future research directions by discussing major components of the generative model.   
\section{Cognitive development and sense-making}

The human mind is an exemplar of a cognitive system exhibiting a high degree of generality in its intelligence. In this section we aim to extract general principles of cognitive development from research areas such as psychology, cognitive science, neuroscience, social psychology, etc.

\subsection{Human cognitive development}\label{sec:human_cognitive_development}

The concept of cognitive development has been defined in the field of psychology as ``the emergence of the ability to understand the world'' \cite[p. 447]{schacter_psychology_2010}. Traditionally it is mostly associated with the child development stages proposed by Jean Piaget but can be also applied to describe sense-making by an individual throughout its whole lifetime as proposed by \cite{kegan_evolving_1982}. Piaget originally contended that children pass through four eras of development - sensimotor, prelogical, concrete operational, and formal operational -  which can be further subdivided into stages and substages \citep{kohlberg_adolescent_1971,piaget_genetic_2004} (see Table  \ref{tab:first_era_of_cognitive_development}). Kegan also propounded that Piaget's and some later cognitive development theories generally describe recursive subject and object relationships when the subject of previous stage becomes an object in the next stage, to which he refers as an `evolution of meaning'. Subject in this context means whatever is perceived as part of self while object is part of environment. Therefore cognitive development can be understood as \textit{an ongoing balancing of subject -- object relations and interactions across the emerging boundary of an individual towards increasing cognitive complexity}\footnote{We formally define the general characteristic of operational complexity of an agent in \ref{subsec:complexity}. Cognitive complexity is an operational complexity \textit{in the context of a cognitive system} characterizing the coupling of internal complexity of a cognitive agent with its environment}. This recursive process progressively defines a boundary of an individual - a psychic differentiation of \textit{self} from the \textit{other} \cite[p. 24]{kegan_evolving_1982} which generally constitutes the differentiation between agent and environment. 

\begin{table}[ht]
	\centering
	\scalebox{.8}{
	\begin{tabular}{| l | p{12cm} | }
\hline
Stage 1. & Reflex action.\\
Stage 2. & Coordination of reflexes and sensorimotor repetition (primary circular reaction).\\
Stage 3. & Activities to make interesting events in the environment reappear (secondary circular reaction).\\
Stage 4. & Means/ ends behavior and search for absent objects.\\
Stage 5. & Experimental search for new means (tertiary circular reaction). \\
Stage 6. & Use of imagery in insightful invention of new means and in recall of absent objects and events.\\
\hline
\end{tabular}
    }
  	\caption{Era I (age 0-2): The era of sensorimotor intelligence. Adapted from \cite[p. 1063]{kohlberg_adolescent_1971}}
  		\label{tab:first_era_of_cognitive_development}
\end{table}

For further clarification of our understanding of cognitive development as individuation and the benefits of such an approach, let us examine a schema of Era I of early cognitive development as formulated by Piaget (Table \ref{tab:first_era_of_cognitive_development}). It is clear that every subsequent stage builds upon the previous one and together they seem to form a hierarchy. It seems however that cognitive development theorists and practitioners, including Piaget, agree that stages in cognitive development overlap, occur in parallel or get manifested later in the maturation process. Therefore we can approach the process of cognitive development as both a sequence of stages and a continuum. In the next section we will see that a developmental continuum punctuated by distinct stages is also supported by understanding cognitive development as a case of individuation. The appearance of stages of cognitive development seems to be better understood in terms of products of individuation or `evolutionary truces' as Kegan calls them.

\subsection{Enaction}\label{sec:enaction}
The enactive approach treats cognition as the adaptive process of interaction between an agent and its environment. The distinction between agent and environment is constituted by the interactions themselves. We define a \textit{cognitive system} as a complex adaptive system which is an organized network of interactive sub-processes \cite[p. 3]{de_jaegher_participatory_2007} that together realize a network of objects and their relations as they are perceived in the world. 

A cognitive system cannot form itself separately from the matrix of interactions with other entities within a larger population. In terms of social psychology this principle is informed by a perspective that minds exist only as social products \cite[p. 328]{summers_object_1994}. Relationships and bonds with other entities of the population are part of the cognitive system and thus define its identity on equal terms with internal relationships and structures. Therefore, the mental states of an individual are not established prior to the interaction, but are shaped, or even created, during its dynamics. \cite{di_paolo_interactive_2012} describe these dynamics as participatory sense-making and propose what they call the `Interactive Brain Hypothesis' which ``describes an extreme possibility, namely that all social brain mechanisms depend on interactive elements either developmentally or in the present, even in situations where there is no interaction'' \cite[p. 5]{di_paolo_interactive_2012}.

Also in some forms of psychotherapeutic theory and practice (e.g. Gestalt, the interpersonal approach to psychoanalysis), certain interactions or situations which are normally considered external to an individual are actually an integral part of its sense-making processes. An individual enacts itself in its social milieu rather than merely using internal representations, plans or theories of mind or even perceptual routines existing prior to the interaction.

\cite{edelman_mindful_1982} define `world inputs' and `self-inputs' to differentiate  between interactions across and within the boundary of a neuronal group. We extend this principle from the context of neuronal groups to networks of cognitive agents. An individual is defined as a totality of both types of interactions while the proportions of them may differ at different periods.

\subsection{Sense-making}\label{sec:emotions_and_sensemaking}
Sense-making is one of the components of the enactive approach to mind and cognition \cite[p. 3]{de_jaegher_participatory_2007}.  We understand cognition as a process of individuation within a scope referred to by \citet{piaget_genetic_2004} as `genetic epistemology'. A psychologically oriented definition of sense-making is: \textit{sensemaking is a motivated, continuous effort to understand connections (which can be among people, places, and events) in order to anticipate their trajectories and act effectively in relation to them} \cite[p. 3]{klein_1._2006}. From the perspective of dynamics of the cognitive system as defined by us, sense-making is continuous effort to form a network of connections and objects as they are perceived in the world. The enactive approach implies that cognition and sense-making are seen not as something that happens inside clearly defined boundaries of the cognitive system but are the product of interactions \cite[p. 1]{mcgann_enactive_2008} across emerging boundaries: ``Sense-making establishes a perspective on the world with its own normativity, which is a counterpart of the agent being a center of activity in the world'' \cite[p. 4]{de_jaegher_participatory_2007}.

In the context of cognitive development, sense-making has the following notable aspects:
 \begin{itemize}
  	\item \textbf{Identity and identification}. A prior notion of an entity `which is making sense' seems to be needed, but in our framework it is not the case: the identity of cognitive agents is created during the process.
	\item \textbf{Enaction}. According to \cite[p. 6]{clark_whatever_2012} perception is an action where an agent produces a stream of expectations and then corrects its own model according to incoming information. Therefore the primary component of sense-making is an action: an agent acts upon the environment, catches the `reflection' or response and updates the internal representation of it.  
	\item \textbf{Reflexive}. Sense-making is a two-way interaction between the individual and its environment across the boundary being created during the same process: any agents' examination, modeling and action `bends' the environment and affects the perception of and further decisions by those same agents. The property of reflexivity of the system captures these mutual influences of networks of processes across the boundary of an agent.
	\item \textbf{Participatory aspect}. As noted by De Jaegher and Di Paolo, ``mental states that `do' the understanding and the ones to be understood are not fully independent or established, but are instead affected, negotiated, and even created as a result of interaction dynamics'' \cite[p. 4]{di_paolo_interactive_2012}. They describe the set of possibilities arising from these dynamics with the notion of \textit{participatory sense-making}, emphasizing its social aspect. In section \ref{sec:conceptual_model} we extend the social aspect of sense-making across multiple scales (Figure \ref{fig:illustration_of_scales}).
\end{itemize}

\subsection{Cognitive dissonance}\label{sec:cognitive_dissonance}

The approach to cognitive development as a sequence of integration and disintegration cycles of meaning is supported by several theories. Leon Festinger's theory of cognitive dissonance, developed in the 1950s, focuses on a state of mind holding two or more elements of knowledge which are relevant but inconsistent with each other \citep{harmon-jones_cognitive_2012}. It is arguably a normal state of an intelligent agent engaged in a life-long activity of making sense of its environment. The theory proposes that incompatibility of the elements create a state of discomfort or `dissonance' which is proportional to the degree of incompatibility - the lack of integration. Further Festinger hypothesized that persons experience an arousal - usually unpleasant emotions - due to the dissonance which motivates them to engage in `psychological work' to reduce the inconsistency. Cognitive dissonance theory in its original form generally enjoys experimental support. Particularly interesting are experiments showing that during the state of dissonance individuals evidence arousal and report negative affect \cite[p. 2]{harmon-jones_cognitive_2012}. Studies in cognitive neuroscience indicate a tendency of a cognitive system to choose a single explanation of sensory experience by constraining multiple possibilities, thereby reducing internal uncertainty or dissonance. For example, the entropic brain hypothesis of \cite{carhart-harris_entropic_2014} points to the association between perception of identity and organized brain activity. The dynamic core hypothesis of \cite{edelman_universe_2000} likewise connects concepts of immediate consciousness with synchronized activity of neuronal groups and areas in the neocortex. 

These observations fit the sense-making concept indicating a tendency of the cognitive system towards increased coherency both internally and in its relationships with the environment. Nevertheless periods of reduced coherency are necessary for the cognitive system in order to explore possibilities of the higher coherency -- what we call \textit{cognitive complexity} in Figure \ref{fig:cognitive_development}.

\subsection{Arousal and emotion}\label{sec:arousal_and_emotion}

Contrary to the established scientific opinion of the end of 20\textsuperscript{th} century, feelings and emotions are just as cognitive as any other percepts \cite[p. 16]{damasio_descartes_2008} and their role cannot be overlooked when considering the development of a cognitive system. While currently the importance of emotions and feelings for the overall operation of a cognitive system is increasingly accepted, the integration of an `emotional system' into the model of cognition is still problematic. \citet[p. 284]{damasio_descartes_2008} proposes a view of emotions as an immense collection of changes occurring in both brain and body, usually prompted by particular content while feeling as the conscious perception of those changes. This proposal is strikingly similar to the two-factor theory of emotion by Schachter and Singer conceptualizing emotion as general arousal plus a cognitive label attached to it \cite[p. 58]{cooper_cognitive_2007}. The state of arousal starts a chain of events within an organism which usually leads to the decrease of arousal. These events can take a form of internal `psychological work' \citep{harmon-jones_cognitive_2012} or external actions in the environment, both of which can be considered as sense-making activities. Further, \citep{damasio_descartes_2008} differentiates between \textit{primary emotions} and \textit{secondary emotions}. Primary emotions are `wired from birth' and constitute what is understood as drives and instincts. Secondary emotions are acquired by creating systematic connections between primary emotions and categories of objects and situations \cite[p. 151]{damasio_descartes_2008}. 

\subsection{The scheme of cognitive development}\label{sec:schema_of_development}

Based on concepts discussed in this section we propose a scheme which conceptualizes cognitive development as an observable sequence of integration and disintegration processes progressively determining the cognitive complexity of an agent. The progressive nature of cognitive development is manifested by \textit{increasing the capacity of sense-making}. This process does not follow a trajectory of monotonous adaptation but rather advances in a punctuated manner going through relatively stable stages. The enactive nature of sense-making implies a reflexive relation between system and environment. At every state, both the cognitive system and the environment have more than one option to relate to each other. Therefore every state of the interaction is characterized by a unique trade-off between freedom and constraints in choosing future trajectory development. Additionally, system--environment boundaries are themselves subject to variation. We suggest that this freedom--constraint trade-off in humans is closely associated with the level of experienced cognitive dissonance. The system achieves higher levels of cognitive complexity via periodic fluctuations in its level of cognitive dissonance. When the cognitive dissonance of the system is low, it undergoes constrained periods of development with more predictable developmental trajectory. When cognitive dissonance is high, the future trajectory of the system becomes more divergent. Our hypothesis is that emotions are mechanisms that guide the selection of the developmental trajectories of the cognitive system by modulating the sensitivity of the system to environmental stimuli. We generalize these mechanisms in our synthetic cognitive development model by introducing value systems that modulate the global developmental activity.

\begin{figure}[ht]
	\centering
	\scalebox{.52}{
	\input{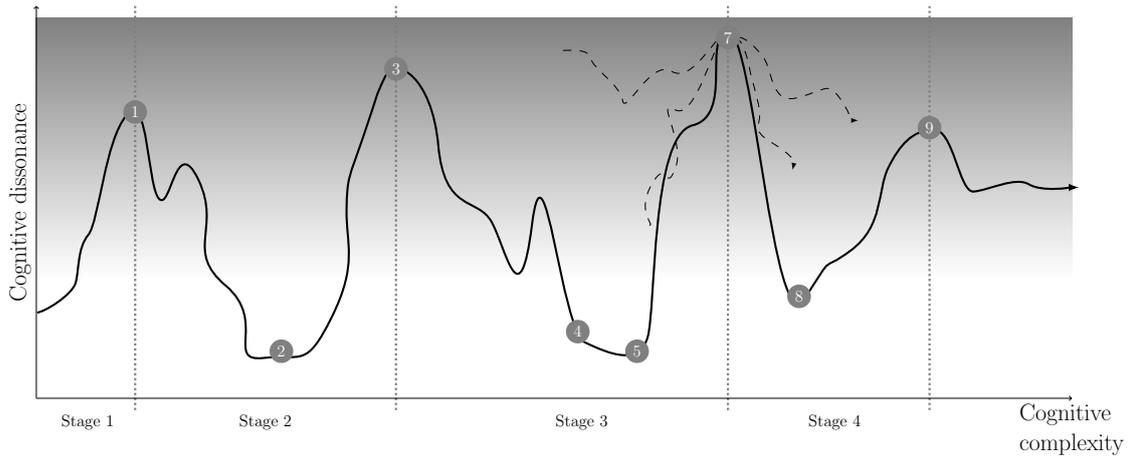}
	}
  	\caption{\textbf{A scheme of cognitive development} qualitatively visualizing the dependency of increasing cognitive complexity on the variation in the level of cognitive dissonance of a system. The bold curve represents actual developmental trajectory. Circles with numbers represent states of development, arbitrarily chosen for illustration. States (1), (3), (7) and (9) mark high cognitive dissonance states where the system has the highest possibility of `choice' between alternative developmental trajectories. Dashed lines are drawn at stage (7) to illustrate multiple possible trajectories that are actually present at every point along the developmental trajectory. States (2), (4), (5) and (8) mark stable periods when the operation of a cognitive system is constrained. Stages 1, 2, 3 and 4 on the horizontal axis illustrate cognitive development stages as described by the developmental psychology representing punctuated manner of increase in cognitive complexity.}
  		\label{fig:cognitive_development}
\end{figure}

Figure \ref{fig:cognitive_development} is a scheme of cognitive development as a variation of cognitive dissonance versus coherence of a system which can be mapped to certain cycles. These cycles emerge from the attempt to balance opposing tendencies to suppress the unpredictability of the cognitive system on the one hand and keep it open to change on the other. 

Human cognitive development is usually understood as a predictable and finite sequence of developmental stages. We argue that both the predictability and finiteness of cognitive development are not ingrained or necessary properties of the process but rather constitute historically shaped superficial characteristics. For example, the relative stability of observable stages of child development are related to more or less stable external influences of parents, peers and society as well as to genetic predispositions. Likewise, the fact that mature individuals rarely undergo transitions to higher levels of cognitive development is possibly related to reduced environmental pressures to engage in the `psychological work' involved. The rationale of seeing cognitive development beyond its observable predictability and finiteness is instrumental for the framework of synthetic cognitive development which aims to describe the genesis of a general cognitive agency as a continuous individuation process, which is the focus of the following section.
\section{Cognitive development and individuation} \label{sec:individuation}
Following the embodied-enactive approach in its broadest sense, cognition is the bringing forth of a world of objects and entities and the relations among them. Bringing forth a world is first and foremost about interacting in it. Perception, action, thought and other cognitive activities are only aspects of this all-encompassing interaction. We can say that cognition, therefore is the bringing forth of individuals, both subjects and objects, and their relations. What an individual is and how individuals come into existence are questions that we address in the context of understanding cognition and cognitive development. As a starting point, an individual is known by that property(ies) or quality(ies) by virtue of which it is unique or describable as such. What define individuals therefore, are distinctions and boundaries the formation of which is a primal activity that precedes even the notion of individual. The nature of distinctions and boundaries is subtle; inasmuch as they separate subject from object, figure from background and one individual from another, they must also connect that which they separate. A boundary, therefore, is not only known by the separation it establishes but also by the interactions and relations it facilitates. 

Following the premise that cognition is the bringing forth of individuals, our interest in cognitive development led us to the interesting conjecture that cognitive development can be understood in terms of how individuals come into existence in the first place, or in other words, to their individuation. As we aim to establish principles of synthetic cognitive development, we find this shift in focus from individuals to individuation; from given identities to their generative processes, both necessary and philosophically profound. 

\subsection{Individuation - a brief philosophical introduction}
To have a preliminary grasp of the concept of individuation, we need to briefly review the importance of individuals in the way we describe the world. The philosophical tradition that started in ancient Greece, and particularly with Aristotle's metaphysics, sees the world as made of individual beings with a given stable identity. What defines an individual is a set of stable qualities. The principle of the excluded middle posited by Aristotle ensures that an individual cannot possess a certain property while simultaneously not possess it. Hence, the identity of individuals, according to Aristotelian theory, is unambiguously defined. To account for the genesis of individuals, according to this theory, we need to identify a principle(s) and the specific initial conditions of its operation that together bring forth the individual. For example, planet earth is an individual object. To account for its individuation, astrophysicists have come up with a theory about the formation of planets and the necessary conditions for planets to form, i.e. the existence of a star such as the solar system. Inasmuch as this scheme makes sense, it suffers a major weakness: it only shows how individuals (planets) are formed from other individuals, i.e. certain necessary conditions that are given \textit{a priori} and an individual principle -- a stable theory of planets formation -- being followed. Clearly, in the very way we commonly think, individuals are primary ontological elements and individuation is only secondary \citep{weinbaum_complexity_2014}. From the Aristotelian theory of individuals, it follows therefore that we must always assume a fully formed individual prior to any individuation. This theory, however, is not very helpful for generalization of our scheme of cognitive development.

Gilbert Simondon was the first to criticize in depth the classical treatment of individuation and the majority of his writings \citep{simondon_individuation_2005} are dedicated to developing a new philosophy of individuation. In \citep{simondon_position_2009} he explains: 
\begin{quotation}	
 ``Individuation has not been able to be adequately thought and described because previously only one form of equilibrium was known--stable equilibrium. Metastable equilibrium was not known; being was implicitly supposed to be in a state of  stable equilibrium. However, stable equilibrium excludes becoming, because it corresponds to the lowest possible level of  potential energy; it is the equilibrium that is reached in a system when all of  the possible transformations have been realized and no more force exists. All the potentials have been actualized, and the system having reached its lowest energy level can no longer transform itself. Antiquity knew only instability and stability, movement and rest; they had no clear and objective idea of  metastability.''
\end{quotation}
In Simondon's new theory of individuation, we are encouraged to understand the individual from the perspective of the process of individuation rather than the other way around \citep{simondon_genesis_1992}. The individual is a metastable phase in a process and is always in possession of not yet actualized and not yet known potentialities of being. Simondon adds:
\begin{quotation}
	``Individuation must therefore be thought of as a partial and relative resolution manifested in a system that contains latent potentials and harbors a certain incompatibility within itself, an incompatibility due at once to forces in tension as well as to the impossibility of interaction between terms of extremely disparate dimensions.'' 
\end{quotation}
According to Simondon, an individual is not anymore a rigid unity with ultimately given properties but rather a plastic entity in a metastable state punctuated by events of transformation. Every such event reconfigures the system of tensions and the manner by which they will determine further transformations. This description is aligned with our understanding of cognitive development as the continuous resolution of cognitive dissonance (see \ref{sec:schema_of_development}).

\subsection{The preindividual}
In their process of individuation, individuals are not preceded by already individuated entities or principles that instruct the trajectory of their formation but by a state of affairs which is yet undetermined -- the \textit{preindividual}. Even after an individual has reached a relatively stable state, the preindividual is not exhausted and persists in the individual. This is what allows its subsequent individuation or becoming. The unity characteristic of fully individuated beings (i.e. identities) which warrants the application of the principle of the excluded middle, cannot be applied to the preindividual. The preindividual goes beyond unity and identity. Deleuze, whose seminal work \textit{Difference and Repetition} draws on many of Simondon's insights, would later describe the preindividual as  ``determinable but not yet determined'' and individuation basically proceeds as its ``progressive determination''\citep{deleuze_difference_1994, weinbaum_complexity_2014}.

Simondon also emphasizes that relations between individuals undergo individuation too: ``A relation does not spring up between two terms that are already separate individuals, rather, it is an aspect of the \textit{internal resonance of a system of individuation}. It forms a part of a wider system.'' \citep[p. 8]{simondon_position_2009} In particular, individuation never brings to light an individual in a vacuum but rather an individual-milieu dyad. This dyad contains both a system of distinctions and a system of relations. The individual and its milieu reciprocally determine each other as they develop as a system wider than the individual.

\subsection{Metastability}
Understanding Simondon's concept of the preindividual -- the dynamic situation that both precedes and is immanent to individuals -- is based on his particular notion of metastability. In systems dynamics, the system's states can be mapped into an energy plane where each state is represented by a point on a N-dimensional plane and is assigned a scalar number designating the energy of the system at that state. A stable state of the system is a state characterized by low energy value relative to neighboring states. If the system is perturbed from a state of stability it will often (depending on the size of perturbation) reach a state of slightly higher energy and will tend to immediately return to the initial state of lower energy. A metastable system is normally a system with a few local minima. Given strong enough perturbations a metastable system may move among states of local stability and hence the designation that implies that no single state is truly stable. It is easy to notice that the topography of the energy landscape here is given and the system dynamics only moves among the already determined set of stable states. Clearly, this representation will only fit an already individuated system. Simondon's notion of metastability departs from this scheme in that the relations between variables in a preindividual condition are not yet determined and the whole landscape is dynamic. As the individuating system moves from state to state, the topography of the landscape changes and may settle into a stable shape only as the state variables mutually determine their relations\footnote{This settlement is also mentioned in an above quote as the resolution of incompatibilities.}. Importantly, the dynamics of a metastable system is not determined \textit{a priori} but rather individuates along with its structure in a sequence of transitions.

\subsection{Transduction} \label{subsect:Transduction}
One of the most significant innovations in Simondon's theory is the concept of \textit{transduction}. Transduction is the abstract mechanism of individuation, an activity which takes place in the preindividual. Classical logic and procedural descriptions cannot be used to think about individuation, because they require the usage of concepts and relationships among concepts that only apply to the results of  the operation of  individuation \citep[pp. 10]{simondon_position_2009}. Transduction comes to designate, therefore, a new model of thought that is constructed from a genetic (as in `genesis') point of view. \citet{combes_gilbert_2013} writes: ``Simondon `transgresses' the Kantian limits on reason.[...] Such an approach  appears to offer a reinterpretation of the thesis of Parmenides, wherein `The same, itself, is at once thinking and being' [...].'' That thought and being (in the sense of individuals brought forth) are considered the same from the standpoint of the mechanism of individuation, highlights how transduction presents a significant contribution to the philosophical understanding of cognitive development. Cognitive development is a formative process where both subject and object are individuated. This individuation produces knowledge -- a resolution, at least a partial one, of an incompatibility (i.e. unresolved tensions) which preexists subject-object differentiation. For Simondon, the conditions of knowledge and of cognition are not given a priori. We can therefore conceive of cognitive development without any inherent limits.        

Simondons adds on transduction: ``One could, without a doubt, affirm that transduction cannot be presented as a model of  logical procedure having the value of  a proof. Indeed, we do not wish to say that transduction is a logical procedure in the current sense of  the term; it is a mental process, and even more than a process, \textit{it is a functioning of  the mind that discovers} [emphasis added]. This functioning consists of  following being in its genesis, in carrying out the genesis of  thought at the same time as the genesis of the object. '' \citep[pp. 11]{simondon_position_2009} (see also p. \pageref{itm:Transdlogic} below).

It is beyond the scope of this article to provide an in-depth review of transduction -- especially of its psychic and social dimensions -- but we will give here the highlights that are essential to our approach to cognitive development. Simondon initially defines the concept as follows:

\begin{quotation}
	``By transduction we mean an operation--physical, biological, mental, social--by which an activity propagates itself from one element to the next, within a given domain, and founds this propagation on a structuration of the domain that is realized from place to place: each area of  the constituted structure serves as the principle and the model for the next area, as a primer for its constitution, to the extent that the modification expands progressively at the same time as the structuring operation. A crystal that, from a very small seed, grows and expands in all directions in its supersaturated mother liquid provides the most simple image of the transductive operation: each already constituted molecular layer serves as an organizing basis for the layer currently being formed.'' \citep[pp. 11]{simondon_position_2009}	
\end{quotation}
Here are a few points that can be extracted from this definition:
\begin{description}
	\item[The~dynamics~of~transduction] Clearly, transduction is reminiscent of the concept of self-organization both in its reference to stability and metastability and in the emergence of structure in a process of relaxing a system of tensions\footnote{We use here tension (or intensity) as a general term for energetic differences that drive structural and state changes in a system. See \citet[chap 2]{delanda_intensive_2013}.}. Individuation will take place as long as the system has not reached a final stability and exhausted its potential for change. But in fact final stability is only an idealization because it requires a closed system that does not interact with its environment, or is  not distinct from its environment (thermodynamic equilibrium). Open systems that maintain at least some distance from equilibrium, or are far from equilibrium, like living organisms and ecosystems, can be considered as continuously individuating. Transduction, however, goes further than the formal understanding of self-organization in addressing complex situations that are more difficult to represent in terms of energy or information exchanges. While self-organization commonly describes the convergence of trajectories towards attractors within an already configured state-space, transduction does not presume such an \textit{a priori} configuration that is characteristic only of already individuated systems. 
	
	Transduction is said to take place when two systems which are initially incompatible come to interact with each other. Simondon uses two terms to explain incompatibility or unresolved tensions: the first is \textit{disparity} which refers to two elements that initially do not share any common ground. The second term is \textit{problematic} in the sense that two systems pose a problem for each other that needs some resolution. For example: the problem the environment poses for an organism which requires it to either adapt or change the environment. The resulting interaction is a transductive process that drives the individuation of both organism and environment. In the course of their interactions, the outcome of which is initially undetermined, they form certain relations of resonance or reciprocal determination by one constraining the dynamics of the other. When such process achieves a relative stability an organization or a structural pattern emerges as the individual. Both initial systems have changed and they now present a pattern of (more or less) regulated interaction that also highlights a distinct boundary. This resolution however is never complete. The remaining unresolved aspects of the interaction are those that maintain the preindividual being within the formed individual and will eventually drive further individuation.          
	\item[Structure~and~operation] Perhaps the most important aspect revealed in transduction is the progressive co-determination of structure and function (see also above regarding the notion of metastability). Individuation can be seen as a chain of operations $O$ on structures $S$: $S \rightarrow O \rightarrow S \rightarrow O \rightarrow S$... \cite[p. 14-15]{combes_gilbert_2013}. Every operation is a conversion of one structure into another, while every structure mediates between one operation and another. Each structure in the chain constrains the operations that can immediately follow. Each operation can transform the previous structure into a limited number of new structures. Every intermediate structure is a partial resolution of incompatibility but it is driven away from its relative stability as long as the existing tensions are not exhausted. This is reminiscent of the propagation of a computation in evolutionary programming. Executed code and the data are analogous to operation and structure. But the code itself is also data that can be progressively modified to produce inexhaustible variety and innovation. This analogy helps us understand how operation and structure are reciprocally determining expressions of the transductive process. 
	\item[Transduction~and~information] \label{itm:information} Simondon's concept of information is substantially different from that of Shannon's theory of information \citep{shannon_mathematical_2001}. Shannon's model of communication seeks to reproduce ``at one point either exactly or approximately a message selected at another point.'' For this, one must presuppose an agreed upon system of encoded messages and the means of their exchange that is already individuated. Simondon's information precedes the individual. He seeks to describe information itself as existing in a state of metastability and indeterminacy. According to him ``information must never be reduced to signals'' but instead, must express the compatibility of two disparate realms \citep{iliadis_informational_2013}. The partial compatibility or coherence that is achieved during transduction is expressed as the emergence of information in the sense of mediating forms or operations. In Simondon's words: 
	\begin{quotation}
	``Information is therefore a primer for individuation; it is a demand for individuation, for the passage from a metastable system to a stable system; it is never a given thing. There is no unity and no identity of  information, because information is not a term; (...) Information can only be inherent to a problematic; it is that by which the incompatibility of  the non-resolved system becomes an organizing dimension in the resolution;(...) Information is the formula of  individuation, a formula that cannot exist prior to this individuation. An information can be said to always be in the present, current, because it is the direction [sens] according to which a system individuates itself.''	\citep[pp. 10]{simondon_position_2009}
	\end{quotation}
	In the context of cognitive development, the individuation of objects necessarily entails the individuation of exchanged signals and their signification across different scales of organization. 	
	\item[Transduction~and~logic] \label{itm:Transdlogic}To further highlight the ontogenetic characteristic of transduction, Simondon compares the process to the logical operations of both deduction and induction. Transduction is not deductive since it does not posit a given principle(s) or pattern(s) external to the process that can instruct the resolution of the present situation. Deduction can only highlight that which is already given by fully individuated knowledge. Transduction discovers relations that did not exist before. Furthermore, transduction is not inductive in the sense that it does not extract those aspects of the incompatible terms that are nevertheless common to them, thereby eliminating what is unique to them. Instead, ``[T]ransduction is, on the contrary, a discovery of  dimensions of  which the system puts into communication [...] each of  its terms, and in such a way that the complete reality of each of the terms of the domain can come to order itself  without loss, without reduction, in the newly discovered structures.'' \citep[pp. 12] {simondon_position_2009} Transduction, therefore, is a real resolution of difference through mediation and not reduction to some common denominator. Following this understanding, we argue that cognitive development is beyond what is reachable by a deterministic logical process. Only after we make sense of something can it become available to the faculties of rational thinking.  
\end{description}
\section{Cognitive development in information-theoretic terms}\label{sec:conceptual_model}

To substantiate our approach to synthetic cognitive development as a process of individuation, the goal of this section is to describe a conceptual model of self-organized boundary formation. We use an information theoretic approach to formalize the dynamic emergence of agents. For the sake of clarity of description, some of the terms here are didactically described from the perspective of an external observer. For example, we describe a distinction between an agent and its environment, or assume distinct states in the operation of agents prior to specifying how such distinctions arise in an actual process of individuation. Such  descriptions are not instrumental to the actual processes; they are given only as a necessary descriptive scaffolding.    

\subsection{Concepts and terms}
\subsubsection*{Agents and boundaries}\label{subsec:preliminary_assumptions}

Let us first consider a heterogeneous population $P$ of $N$ agents that are capable of exchanging information via interconnections. Each agent has a set of input and output connections to other agents so that they together form a network which reflects a topology of information exchanges. For simplicity we assume for the moment that each agent realizes a state machine with an internal state and input and output vectors. Input information is processed depending on the current internal state to produce an output and possibly change the internal state of the agent. We also assume that all the agents in the population are relatively stable in their behavior. As will become clearer in the following, the state machine description of the agent is only a schematic representation and does not reflect the actual structure of the agent.

In the course of individuation, $P$ is differentiated by self-organizing into two subsets $U$ and $E$. The subset $U$ operates in relation to subset $E$ as an environment that provides a repertoire of changing signals. Some of the agents in $U$ can sense changes in the environment $E$ (sensors) while others can introduce effects in the environment (actuators). As we consider self-organized boundary formation, prior to such formation, $P$ is a pre-individual undifferentiated population. The division of $P$ into $U$ and $E$ is in fact the product of an actual individuation process which will be specified by a complementary generative model. Here we only specify the initial and final stages of the process. 

\subsubsection*{Heterogeneity and Redundancy of \bm{$P$}}
The agents of population $P$ are heterogeneous and redundant. Each agent in the population receives input signals and produces output signals that depend on its inputs and internal state. The transfer functions (i.e. input/output relations) of agents form a continuum of behaviors which account for the heterogeneity of the population. In addition to the variety of behaviors, there is an additional variety in the way each behavior is realized by agents with similar transfer functions. This accounts for the population's redundancy. Seen by an external observer, the behaviors of agents can be loosely classified into various functions such as sensors, actuators, filters, integrators, logical gates etc. But as our model focuses on individuation, the particular transfer functions do not matter. What matters is the generative process that organizes the population or parts of it into a coordinated whole where distinctions and coherent relations emerge. Both the heterogeneity and redundancy of the population are essential to the selective mechanisms that drive the process of individuation and will be further discussed in section \ref{sec:towards_gen}.

\subsubsection*{Individuation}
Our goal is to model cognitive development as the process of individuation described in section \ref{sec:individuation}. Specifically, we aim to model the process by which a subset $U$ of $P$ self-organizes and differentiates itself from $E=P-U$ thus giving rise to a boundary and an individual--milieu distinction. The outcome of this process is a new individual agent\footnote{In the following it will become clear that more than one integrated agent can emerge in the process.} $A$ whose function is realized by the coordinated and synchronized operations of its constituent agents belonging to $U$ that are reciprocally selected from $P$. The term reciprocally selected here means that agents in the population spontaneously select each other, as they interact, to form coordinated coalitions without the intervention of an external guiding agency. As already mentioned in section \ref{sec:individuation}, the produced individuals are never final products. Changes in the environment may bring about changes in the structure and function of the emergent agent. Moreover, every such agent $A$, once emerging, may disintegrate altogether. That is why the stability of the structure and function of agents in our model is only a relative stability. In fact, all agents are dynamic constructs.

\subsubsection*{Scales of individuation}\label{subsec:scales_of_individuation}
In our model, individuation is a scalable process that takes place at multiple scales, both structural and functional, of the individuating system. We describe the model at some scale $S$, where we observe a population of agents $P_{s}$. Every agent in $P_{s}$ is a product of self-organization of simpler agents at the lower scale $S-1$. Similarly, super-agents $A_{s}^{i}$ that emerge at scale $S$ are the elements at the higher scale $S+1$. The individuation of agents, therefore, is taking place simultaneously at multiple scales. In most cases, lower scale agents must have more stable properties than higher scale agents. Instability of agents at lower scales would make higher level organization much less probable.

Scales differ not only structurally but also temporally. As the cognitive system individuates, complex objects emerge and their frequency of interactions become slower in comparison to their lower scale components. Generally, therefore, the relative frequency of interactions at scale $S$ is slower than the rate of change at scales lower than $S$ and faster than the frequency of interactions at scales higher than $S$. It is helpful, therefore, to understand the time scale associated with population $P_{s}$ as the average duration of interactions within the population.

\begin{figure}[H]
	\centering
	\scalebox{.23}{
	\includegraphics{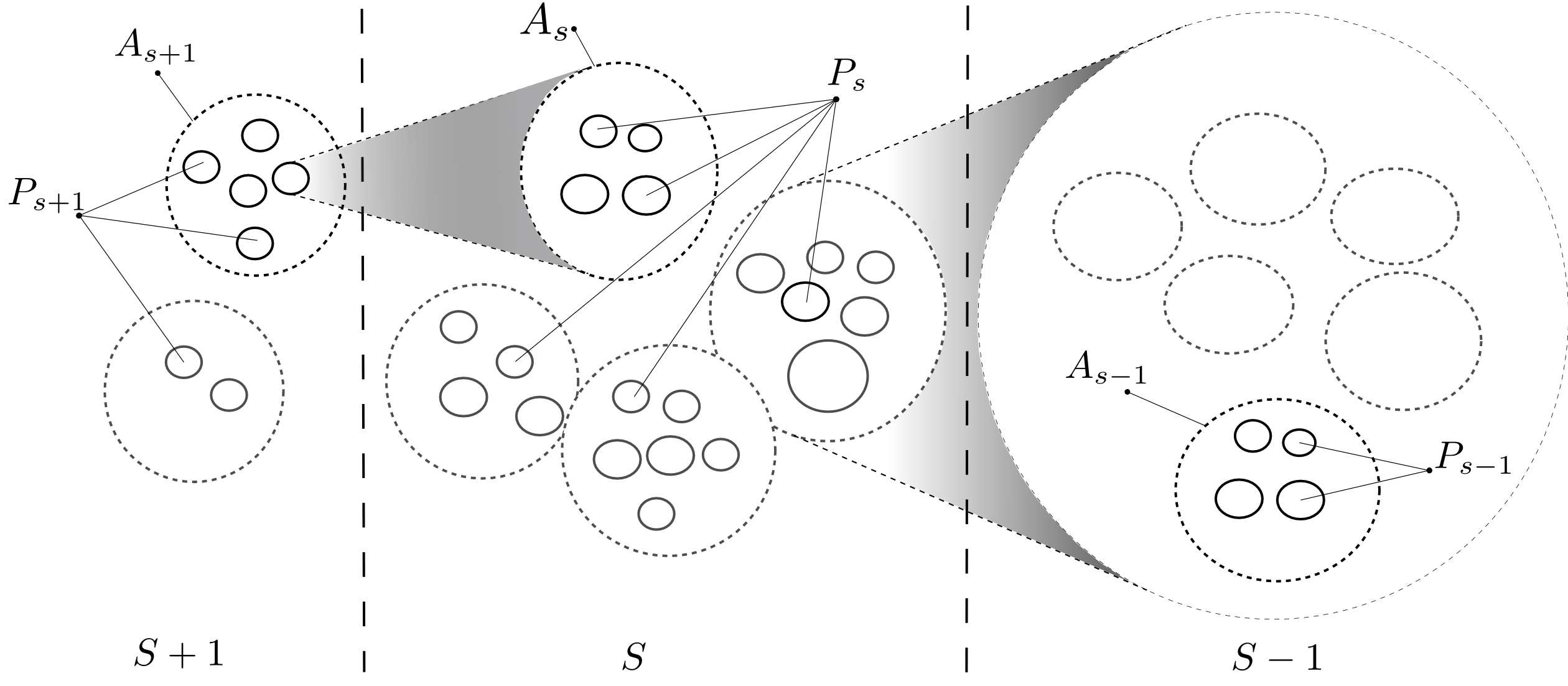}
	}
  	\caption{Relationship among scales, populations and boundaries in the model. The chosen scale of analysis is $S$. $S+1$ is the higher scale while $S-1$ is the lower scale. $P_{s}$ denotes a population of agents at scale $S$. Solid circles denote the agents of population $P$ at any scale. Dashed lined circles denote super-agents at any scale e.g. -- $A_{s}$ at the center of the figure, denotes a super-agent that emerges from the interactions of agents in $P_{s}$. Super-agents at scale $S$ are the agents of the population $P_{s+1}$. The $i-th$ super-agent at scale $S$ is denoted $A_{s}^{i}$, the superscript is omitted if unneeded. Also, the subscript $S$ is omitted from $A$ or $P$ in the text if it is redundant.}
  	\label{fig:illustration_of_scales}
\end{figure} 

Following Simondon's understanding of information (see \ref{itm:information}), as new individuals $A_{s}^i$ emerge at scale $S$, new information is being created. This information is expressed in the structural and functional distinctions that become apparent at that scale. Whatever remains incompatible among the agents of the lower scale does not get expressed in the emergent new structures. Across multiple scales of individuation, these incompatibilities remain as the preindividual in their respective scales. 

As we will see in \ref{subsec:complexity}, the emergence of a new individuated organization at scale $S$ is accounted for in the reduction of entropy at the lower scale. But the new individuation(s) do not correspond to merely the entropy now calculated at a higher scale, but to the internal complexity $C_{intrn}$ expressed in equation \ref{eq:internal_complexity}. The reason behind this difference is that the new individuals at scale $S$ retain some incompatibility which is not resolved. They are  therefore not ultimately organized, thus exhausting any further individuation. $C_{intrn}$ takes into account the actual compatibility (in terms of mutual information) between the constituting agents and not merely the maximal repertoire of states they can present. This `internal coherence' expressed by $C_{intrn}$ can indeed be associated with the amount of preindividual information that turns into a new individuated form.   

\subsection{Definition and formation of boundaries}\label{description_of_boundaries}
In our model, the boundaries defining the agent -- environment distinction and the relations between them -- are never entirely fixed. The functioning of any emergent agent is adaptive and subject to changes. We follow the work of Giulio Tononi and his concept of \textit{information integration} to formally define coordinated clusters in networks of interacting agents \citep{tononi_information_2004,tononi_consciousness_2008,edelman_universe_2000}. The reasoning behind the concept is that if we examine a population $P$ of $p_{i}$ interconnected agents, where $i \in [1,..,N]$, we wish to quantify how much they affect and are being affected by each other. In information theoretic terminology, each agent $p_{i}$ can either change its state independently of all other agents in $P$, or its state may depend on the states of other agents in $P$, or even be entirely determined by the states of other agents. The mutual information between two agents $p_{i},p_{j}$ is given by the formula: 
\begin{align}
	MI(p_{i},p_{j})=H(p_{i})-H(p_{i}/p_{j})=H(p_{j})-H(p_{j}/p_{i})\\
	=H(p_{i})+H(p_{j})-H(p_{i},p_{j})
\end{align}

Where $H(x)$ is the entropy involved in the state of agent x. If $p_{i}$ and $p_{j}$ are independent, $H(p_{i},p_{j})=H(p_{i})+H(p_{j})$ and then $MI(p_{i},p_{j})$ would be 0. The mutual information would be maximum in the case that the state of one agent is fully determined by the other. In this case the mutual information will be equal to $min(H(p_{i}),H(p_{j}))$.

For a set of agents $p_{i}$ in $P$ the integration of the whole set would be given by the sum of the entropies of the independent agents $p_{i}$ minus the entropy of the joint set $P$:
\begin{equation}
	I(P)=\sum_{i=1}^{k}H(p_{i})-H(P)	
\end{equation}

In order to compare the degree of integration within a subset of agents to the integration between the said subset and the rest of the population, we divide the population of agents $P$ into two subgroups of differing sizes: $X_{i}^{k}$ and its complement $P-X_{i}^{k}$, where k is the number of agents in the subset $X$. The mutual information between $X_{i}^{k}$ and its complement is:
\begin{equation} \label{eq:MI}
	MI(X_{i}^{k},P-X_{i}^{k})=H(X_{i}^{k})+H(P-X_{i}^{k})-H(P)
\end{equation}

Formula \ref{eq:MI} measures the statistical dependence between a chosen subset $i$ of $k$ agents and the rest of the population. The \textit{Cluster Index $CI$} of the subset $X_{i}^{k}$ will therefore be given by:
\begin{equation}
	CI(X_{i}^{k})=I(X_{i}^{k})/MI(X_{i}^{k},P-X_{i}^{k})
\end{equation}
$CI$ measures the degree of distinctiveness of a subset of agents in $P$ compared to the whole population in terms of information exchange\footnote{It is important to note that these are only simplified formulas that do not take into account the different sizes of subsets.}. For $CI\le1$ there is no significant distinctiveness while a subset with $CI\gg1$ indicates a distinct integrated cluster.

 Equipped with this comparative measure, we can compute the subset $X^{k_{max}}$ of maximal size $k_{max}<N$ in $P$ with the highest cluster index that \textit{does not include subsets with a higher cluster index}. This subset will be designated as the primary functional cluster $PFC_{p}(t)$ of population $P$ at time $t$. The primary functional cluster corresponds to the super-agent $A$, while the set of its constituent agents corresponds to $U$ (see \ref{subsec:preliminary_assumptions}). The definition of functional clusters makes concrete the differentiation between an agent and its environment. Considering the relevant rate of information exchange in the population $P$, the computation involved can repeat itself in appropriate time intervals to yield a time series of $PFC$s. In other words, the boundaries of the primary functional cluster of $P$ and the  particular agents participating in it vary in time. 
 
 \subsection{The dynamic core}
 The time-dependent primary functional cluster $PFC(t)$ will be the one corresponding to the super-agent $A$ brought forth by $P$. The dynamic entity thus created was termed by \citet[chap. 12]{edelman_universe_2000} the \textit{dynamic core}. However, our usage of the term is not confined to modeling the central nervous system. 
 In the simplified case above we consider that $P$ produces at any given moment only a single super-agent but clearly this is almost never the case. Actually $P$ can give rise to a number of emergent integrated agents. Such agents can be identified by recursively repeating the above computation on $P$ to yield a set of functional clusters $(PFC_{p}(t),FC_{p}^{1}(t),FC_{p}^{2}(t),...,FC_{p}^{m}(t))$\footnote{See \citep[chap. 14]{edelman_universe_2000} for discussion of multiple functional clusters.}. The members of such a set are all distinguishable in terms of their integration relative to their environment as long as their clustering index $CI$ is significantly greater than 1. In the simple case however the $PFC_{p}(t)$ is the individuated agent and the rest of $P$ is considered its environment $E$. We also assume that $PFC_{p}(t)$ is relatively slow-changing in time compared to the rate of information exchange between the agents that constitute it, but still it can present a rich repertoire of states due to both its interactions across the boundary and the interactions of agents inside its boundaries. 

\begin{figure}[H]
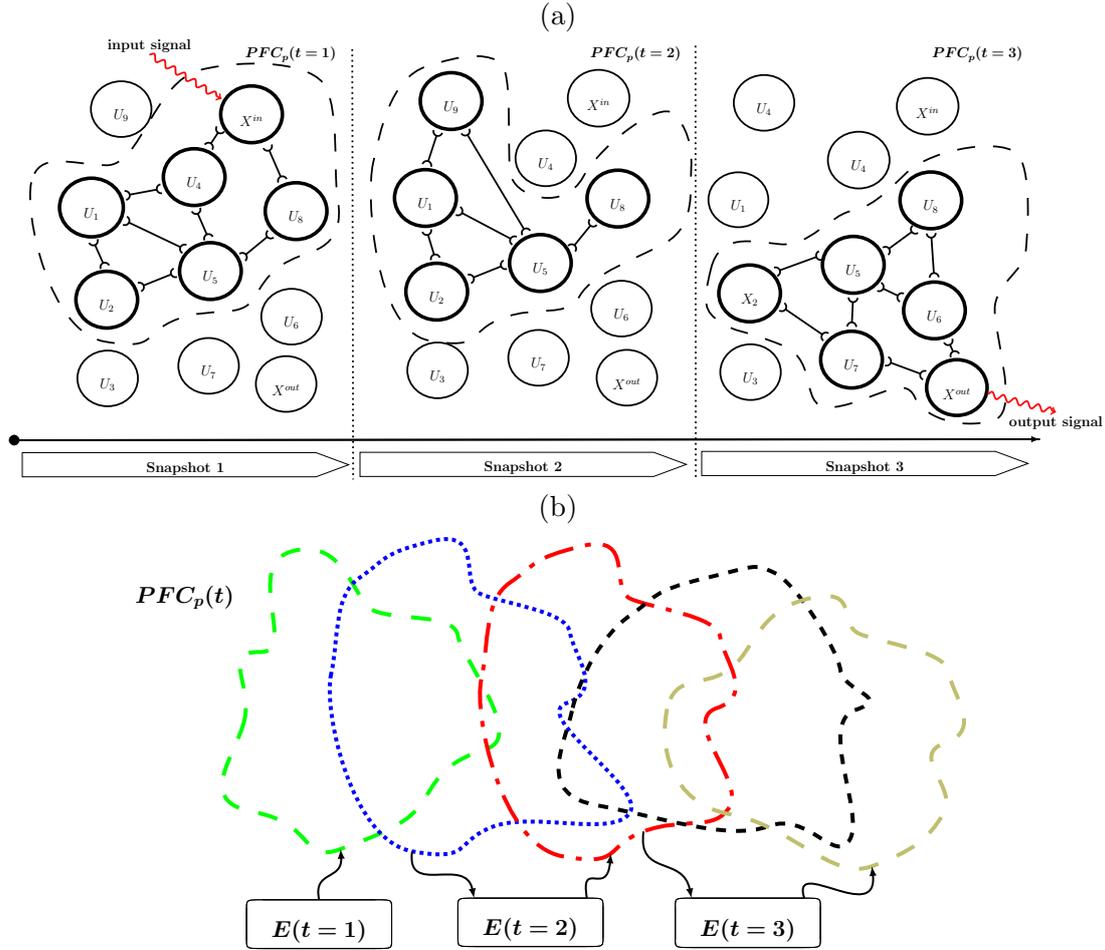

	\centering
	(a)\\
	\scalebox{.45}{
	\input{pictures/functional_clusters.tex}
	}
	(b)\\
	\scalebox{.5}{
	\input{pictures/dynamic_core.tex}
	}
	
  	\caption{Top figure (a): $PFC_{p}(t)$ developing along three consecutive snapshots in time. The boundary of the clusters moves from top right of the population to bottom right. The sequence of primary functional clusters depends on connectivity patterns, which results in the observed drift of the boundary. At the level of super-agent $A$'s interaction with its environment, the given example of the drift of boundary illustrates a non-trivial response of $PFC_{p}(t=3)$ through $X^{out}$ to a stimulus acquired by $PFC_{p}(t=1)$ through $X^{in}$. Bottom figure (b): the dynamic core interacting with its environment during five consecutive snapshots.}
  		\label{fig:functional_cluster}
\end{figure}

A functional cluster (Figure \ref{fig:functional_cluster}a) is a set of agents which exhibit a synchronized activity for a limited duration. Functional clusters reveal integrated structural and functional properties of the network and are \textit{snapshots} $(PFC_{p}(t=1),PFC_{p}(t=2),...)$ of a continuous process of agents exchanging information with each other i.e. the dynamic core. The dynamic core's boundary drift can be observed as the temporal sequence of functional clusters (Figure \ref{fig:functional_cluster}b). Remarkably, the dynamic core is defined in terms of interactions among agents, rather than merely in terms of their topological or functional relations \citep[p. 159]{edelman_universe_2000}.

 \subsection{Operational complexity} \label{subsec:complexity}
 A major characteristic of functional clusters pointed out by Edelman and Tononi \citeyearpar[chap. 11]{edelman_universe_2000} is their complexity. Considering an agent as an integrated functional cluster, the complexity of the agent quantifies how differentiated are the agent's inner states, or, in other words, how many different states it can activate. Complexity is closely related to integration. If all the agents of a cluster had operated independently from each other, the cluster could have been said to have the highest complexity as its entropy would have been maximized. But then, with $CI\ll1$ such a cluster is not a cluster at all but a collection of independent agents. Similarly, at the other extreme, a very high integration means very few possible states of the overall cluster, or, in other words, the states of individual agents of the cluster are mostly determined by global states. In such a case $H(PFC_{p})$ approaches zero. The complexity of functional clusters therefore depends on the entropy of subsets within the cluster and the mutual information among them. Let $X$ be a $PFC$ of size $M$ in the original population $P$. We assume that $X$ is isolated from its environment\footnote{The cluster can be initialized to some arbitrary initial condition.} so its inner states are self produced. We divide $X$ into two complementary subsets $X_{j}^{k}$ and $X-X_{j}^{k}$ of respective sizes $k$ and $M-k$. The index $j$, enumerates all possible subsets of size $k$ out of $X$. The internal complexity of the cluster $X$ can be given by:
\begin{equation} \label{eq:internal_complexity}
	C_{intrn}(X)=\sum_{k=1}^{M/2}<MI(X_{j}^{k},X-X_{j}^{k})>
\end{equation} 
where the mutual information is averaged on all subsets of size $k$. Clearly, subsets of very small size will contribute very little to $C_{intrn}(X)$, while subsets of sizes in the vicinity of $M/2$ will contribute the most complexity. Remarkably, $C_{intrn}$ measure of complexity is based only on the extent to which subsets of the cluster affect each other and the statistical properties of the signals that agents within the cluster exchange. $C_{intrn}$  therefore does not rely on an arbitrary measure of complexity imposed from outside the cluster. 

To complete the picture, we consider two special subsets of the cluster $X$. $X^{in}$ is the subset of the agents that receives signals from the environment across the boundary of the cluster (i.e. sensors), while $X^{out}$ is the subset of agents that transmits signals to the environment across the boundary (i.e. actuators). $X^{in}$ and $X^{out}$ can of course overlap. Two additional useful quantified characteristics of a functional cluster is the degree to which it can be affected by its environment and the degree to which it can affect the environment. If the environment is defined as all the agents in $P$ which do not belong to $X=PFC_{p}$ then we can define the environment as: $E=P-PFC_{p}$. While holding the states of all the agents in $X-X^{in}$ in constant state, we can compute the input complexity as:
\begin{equation}
C_{in}(X)= MI(X^{in},E)
\end{equation}
Similarly, the output complexity can be computed by holding $X-X^{out}$ in a constant state:
\begin{equation}
C_{out}(X)= MI(X^{out},E)	
\end{equation}
Finally, the overall interactive complexity can be given by the mutual information of $PFC_{p}$ and $E$ when nothing is held constant:
\begin{equation}
	C_{io}=MI(X,E)
\end{equation}
These are of course simplified formulas that do not take into account the time dependencies of $X,E,X^{in},X^{out}$ and the variation in the dynamics of $X^{in},X^{out}$ that may arise from the different instantiations of the parts that are held constant in the computation. Nevertheless, even with these simplified formulations, the clustering index $CI$ together with the various complexity measures $C_{intrn},C_{in},C_{out}$ and $C_{io}$ inform us about the general characteristics of the emerging super-agent and its dynamics. $C_{intrn}$, for example, may indicate the memory capacity of the agent; the more integrated internal states are, the more `experience' the agent can potentially draw from in its interactions.       
\section{Towards a generative model} \label{sec:towards_gen}
In this section we outline the foundations of a generative model for cognitive development. By generative model we mean an implementable architecture that will demonstrate scalable cognitive development in terms that we described above, i.e. processes of self-organized differentiation, boundary formation and object relations. We identify two general mechanisms that are essential to our model, namely the transductive mechanism and the value modulating system.

The basic architecture of our cognitive development model is a heterogeneous population of agents that interact via links, thus forming a network. Information exchanges among agents result in the self-organization of clusters of agents that synchronize and coordinate their activities. Such self-organized clusters operate as distinct agents at a higher scale of organization. The most elementary formative processes in our model are the creation, reinforcement, suppression and destruction of links among agents. These formative processes are guided by the nature of exchanges among agents, e.g. Hebbian reinforcement rules in neural nets. All higher level formative mechanisms in our model are constituted from these elementary processes and their modulation.           

\subsection{Transduction}
The mechanisms that are responsible for the formation of boundaries and the bringing forth of coordinated activities in a population of agents $P$ arise primarily from the agents' intrinsic capabilities to affect and be affected by each other. The specific characteristics of the interactions, e.g. their frequency, their synchronization and coherence, have a critical influence on the way agents are connected. Such influence finds its expression in the reinforcement or suppression of connections among agents and consequently on how strongly they may actually affect each other. This is how \textit{the activity of agents within $P$ progressively determines the topological organization of the network of agents in $P$}. The structural organization, in turn, affects the overall function of the individual agents by selecting interactions. This recursive process of activity-determining structure and structure-determining activity is described in \ref{subsect:Transduction} as transduction\footnote{Deleuze uses the term \textit{progressive determination} to describe the same process. See,\citet{weinbaum_complexity_2014}.} -- the driving mechanism of individuation. 

Individuation is described as taking place when two incompatible systems interact and achieve a certain degree of compatibility. In the generative model that we develop, agents in population $P$ interact and mutually select other agents with whom they are compatible.
The connections between compatible agents are reinforced while other connections tend to be suppressed. In the course of such recursive selective interactions, groups of compatible agents cluster into distinct compound organizations resulting in individuated super-agents. What needs to be further clarified is the criteria for compatibility utilized in the selective process and the actual mechanism of reflexive mutual selection taking place among agents.

\subsubsection*{Criteria for compatibility}

Agents overcome their initial incompatibility by constraining each others' regimen of behaviors. In other words, there is a process of reflexive selection going on where every agent selects with which other agents in the population it can interact. We present here three understandings of the concept of compatibility from the simpler to the more complex. A concrete selective criterion of the adaptation of link strengths among agents is derived accordingly from each understanding:   
\begin{description}
	\item [Synchronization] -- Agents that produce effects (become active) at the same time will tend to  reinforce their connections.\footnote{This does not exclude the formation of new links as well.} The kind of compatibility that is selected by this criterion is temporal coincidence, which may indicate with some probability that the synchronized agents are causally affected by either the same event or by events that are causally connected, or events that are otherwise correlated. The formation of synchronized clusters of agents is the simplest form of individuation. Synchronized groups will tend to reinforce their synchronized behaviors and suppress their out-of-sync behaviors. Examples of individuation following this criterion can be found in neural networks. The Hebbian rule that neurons that fire together also wire together is one application of this criterion. A more complex application is provided by \citet{edelman_universe_2000} who hypothesize that spontaneous synchronization among groups of neurons is the basis of consciousness.\footnote{See also \citet{tononi_reentry_1992} for a more detailed description.} From the perspective of our approach, both are examples of cognitive development at the scale of groups of neurons.      
	\item [Coherence] -- Agents that produce effects (become active) in response to informative patterns (not necessarily synchronized) that represent the same category or type, or a group of mutually supporting logical propositions, or a group of associative patterns, will tend to reinforce their connections. The kind of compatibility that is selected by this criterion is much more abstract then synchronization and requires a context  of operation. The agents connecting according to this criterion form coherent clusters. Clearly, in this general form, the coherence criterion is underspecified. Coherence will normally operate as a selective criterion only in populations of relatively complex agents where the information that agents exchange already signify lower level individuated objects. Such objects provide the context that further determines what coherence means. \citet{thagard_coherence_2002} explains coherency as the joint property of propositions that tend to be selected together or rejected together when tested in the context of a certain domain or state of affairs. In our case, Thagard's understanding of coherence distills a second kind of compatibility, which we can generally describe as compatibility in signification or meaning.\footnote{By incorporating various selective criteria, our approach to synthetic cognitive development suggests a unifying ground to both connectionist and symbolic models of intelligence. We see this as a promising prospect for further research.}         
	\item [Coordination] -- Coordination is broadly defined as functional compatibility. In fact, synchronization and coherence can be described as special cases of coordination. Agents that interact, process information and produce effects that jointly realize a function or a goal are said to coordinate their operations, thus presenting functional compatibility. Connections among agents that support the coordinated activities will be reinforced while those that disturb the coordinated activities will be suppressed. The agents connecting according to this criterion will form coordinated clusters. As in coherence, coordination will operate as a selective criterion only in populations of relatively complex agents and where the information that agents exchange already signifies lower level individuated objects and their relations. Such objects provide the context that further determines the nature of the function or goal that are performed by the coordinated clusters. Autopoiesis \citep{maturana_autopoiesis_1980} is perhaps the most illustrative example of a self-organized coordination. Remarkably, autopoeisis is a function that operates in relation to the same cluster of agents that realizes it and therefore does not require an outside observer for its definition. Functional compatibility is not limited to this family of self-determined functions. Coordinated clusters may emerge in response to signals mediated by the value system (see \ref{subsec:value}) that are external to the population of agents under consideration. Such signals guide selection by providing an external criterion of functional efficacy. In other words, the actual compatibility criterion of coordination may be either self-produced or external. The emergent agents, accordingly, may be self-coordinating or coordinated in relation to an external state of affairs. For an overview of coordination mechanisms see \citet{heylighen_self-organization_2013}.        
\end{description}

Though this short list of criteria seem to cover a very wide range of individuating processes it is not necessarily exhaustive. Novel understandings of compatibility may emerge in the course of an open-ended cognitive development process. However, we see no problem over incorporating such future developments into our model. On top of the criteria of compatibility we identify additional criteria under the general title \textit{value modulating system}, or in short, value system. Values are not used explicitly in selection. Instead, they operate by modulating the plasticity of connections and by that quicken or slowdown the selective processes. The value system and its importance is further discussed in \ref{subsec:value}.  

\subsubsection*{Reflexive mutual selection}
In our generative model, individuated entities are the product of a recursive resolution of incompatibilities. This is a process of reciprocal selection where agents within a population repeatedly select communication links and interactions that increase compatibility according to the criteria outlined above. The reinforcement of compatible interactions and suppression of incompatible interactions progressively determine clusters of integrated agents within the population. Structural changes in the network of agents drive further selections and this transductive activity continues until the network achieves relative stability as individuated super-agents consolidate. At this elementary level, individuals emerge as products of an evolutionary process: the heterogeneity of agents in population $P$ provides the variation and the various compatibility criteria provide the selective elements of the process. The retention of compatible clusters is inherent in the process since by definition mutual compatibility among agents is preferred and reinforced. Otherwise, no individuation and no cognitive development could have taken place.      

Inspired by Edelman's theory of neuronal group selection \citep{edelman_neural_1987,edelman_reentry:_2013,tononi_reentry_1992} the reflexive and recursive characteristics lie at the basis of our generative model. Our model extends neuronal group selection to general networks of agents. The selective criteria of compatibility that we derive from the theory of individuation extends the synchronization criterion in the case of neuronal groups. Reflexive mutual selection (termed `reentry' by Edelman) is a mechanism operating within a network of interacting agents. Consider two groups of agents A and B. Each group contains similar agents with some variety in their pattern of behavior. The groups are interconnected internally and across. Following a signal produced by some agent in group A, a subset of agents in group B will respond by producing signals too. This activation will spread both internally in B and across back to A (where some of the agents are already active too). A subset of agents in A will respond to the signals coming from B such that a chain reaction of signals will propagate back and forth between A and B. In some cases, after a few cycles of exchange, a signal, whether from an agent in B or A, will be received by the initiating agent and will cause it to produce a signal similar to the one that initiated the whole exchange. If this happens, a closed activation loop begins and the groups will enter a period of sustained mutual activation that will continue until it is disrupted by other signals.\footnote{The description here is simplified for the purpose of illustration. Sustained activation patterns may arise in many other, more complex ways.} Sustained activation patterns and sequences of interactions that arise in a similar manner within the population of agents are the products of what we call a reflexive mutual selection process.  

\begin{figure}[H]
	\centering
    	\scalebox{.4}{
    \input{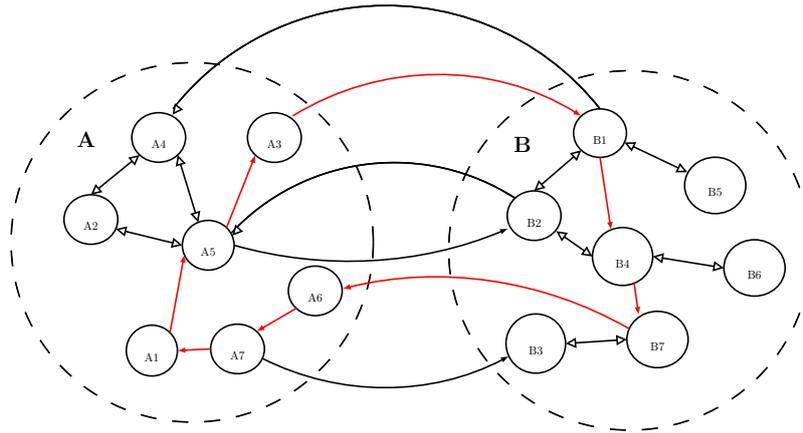}
    }
  	\caption{Two connected groups A and B and the formation of sustained mutual activation. The signal propagation path of the sustained activation is indicated in red. Other paths such as A3-B1-B2-A5-A3 are topologically possible but are not selected because activation depends also on the informational content and timing.} 
  		\label{fig:agents_and_units}
\end{figure} 
\subsection{Value modulating systems} \label{subsec:value}
In the context of the cognitive development of the human mind, value systems hold a potential to describe the mechanism of emotional influence on cognitive states, which is increasingly seen as a pivotal component of cognition. \citet{pessoa_cognition_2009} points out: ``Historically, emotion and cognition have been viewed as largely separate. In the past two decades, however, a growing body of work has pointed to the interdependence between the two''. \citet{fellous_human_2004,fellous_neuromodulatory_1999} argues that neuromodulatory systems are the basis for emotions. In summary, emotional regulation is an important aspect of cognitive development (see \ref{sec:arousal_and_emotion}). We therefore assign to value systems a critical regulatory role in our model. 

In the enactive theory of cognition, \citet{di_paolo_horizons_2010} offer a definition for value as ``the  extent  to which a situation affects the viability of a self-sustaining and precarious network of processes that generates an identity.'' Moreover, they emphasize a strong relation to the concept of sense-making: ``value  is  simply  an  aspect  of  all  sense-making,  as  sense-making is, at its root, the evaluation of the consequences of interaction for the conservation of an identity.'' Values are assigned to situations but they emerge from interactions as aspects of sense-making or, in other words, values emerge in the course of cognitive development. In the context of our research program we aim to examine how values and value systems develop. Particularly, the relations between built-in values and emergent values in the process of individuation at any given scale. 

In the neuroscientific context \citet[p. 2]{friston_value-dependent_1994} take an evolutionary approach:
\begin{quote}
``The value of a global pattern of neuronal responses to a particular environmental situation (stimulus) is reflected in the capacity of that response pattern to increase the likelihood that it will recur in the same context. In this respect, value is analogous to `adaptive fitness' in evolutionary selection, where the adaptive fitness of a phenotype is defined in terms of its propensity to be represented in subsequent generations. Thus, value plays a role in neuronal selection similar to that which adaptive fitness plays in evolutionary selection.''
\end{quote} 

Friston et al. highlight the role of values as mediating evolutionary knowledge significant to fitness and already proven survival strategies. Cognitive development will accordingly be guided by values built-in by evolution. This approach assumes certain abstract principles (e.g. ``food is good'') that are independent of interaction and exist \textit{a priori} to individuation. Di-Paolo et al. argue against this approach as a case of ``[...] dealing with \textit{pre-factum} evolutionary teleonomy, not with autonomy'' and further explain that ``[t]he point is not to argue that such norms do not exist across individuals [...], but rather that they should be  searched for on the emergent level of autonomous interaction, not on the level of mechanism.'' We go a step further to argue that the emergent level mentioned by Di Paolo et al. extends beyond the individual (i.e. the autonomous entity) into the process of individuation.   

But the gap between these approaches may be merely superficial. According to Friston et al., the structural and functional properties of the value systems needed for guiding neuronal selection should: (1) ``be responsive to evolutionary or experientially salient cues'' i.e. a wide context; (2) ``broadcast their responses to wide areas of the brain and release substances that can modulate changes in synaptic strength''; (3) be ``capable of a transient response to sustained input, inasmuch as it is changes in circumstances (environmental or phenotypic) that are important for successful adaptation''. Value systems allow for the integration of broad contextual information in driving selective processes. \citep{friston_value-dependent_1994} note however that a value is equivalent to an adaptive fitness in the evolutionary sense which guides, but \textit{does not determine}, the further development of the organism. 

In our model, we frame value systems in a broader framework of scalable individuation. We accommodate pre-determined norms in the dynamic construction of significance by individuals interacting in their environment. We argue that value modulating systems offer mechanisms of upward and downward causation which mediate among the various scales of the cognitive system (see Figure \ref{fig:illustration_of_scales}). Values at a specific scale $S$ operate as guiding signals originating from both the lower scale $S-1$ and the higher scale $S+1$. In both cases they modulate the  operation of agents within the population $P_{s}$. Values that originate from lower scales can be viewed as built-in but they cannot be said to operate at the same domain of interactions characteristic to scale $S$. Every scale is a new layer of mediation whereby the possibilities to create meaning for signs become less constrained by the values of lower scales \footnote{See \citep[pp. 48-52]{di_paolo_horizons_2010} for a more detailed discussion that supports this approach.}. 

Synthetic cognitive development starts from explicit presuppositions and constraints which provide the basis for the generative process. These presuppositions define the primary repertoire of structures available for further cognitive development e.g. eyes, ears, pain receptors with their related neural structures in the case of mammals. Likewise, our notion of a value system starts from some presuppositions and constraints i.e. `innate values' that provide the initial structures. Value systems, then, undergo individuation along with the whole cognitive system. \citet[p. 10]{friston_value-dependent_1994} demonstrated that value systems allow for unsupervised acquisition of new values in cases where they predict behaviors related to innate ones\footnote{In `innate' we understand whatever is necessary for the viability of the individuals that are brought forth at a given scale. Value signals that originate form higher scales may however destabilize individuals whose continued existence is not significant anymore at higher scales of individuation.}. Following this, we hypothesize that similar mechanisms can be applied generally, meaning that novel sets of values are recursively acquired based on previously established ones. The specific characterization and implementation of such mechanisms is the subject of future research. 

Presuppositions and constraints are necessary for modeling concrete instances of 
the cognitive development process. We nevertheless adhere to the perspective that the general cognitive development scheme based on the theory of individuation is not fundamentally constrained by any particular set of such presuppositions. Individuation as a formative process spans across scales beginning with natural evolution and going as far as open-ended intelligence expansion in humans, human organizations and machines.

\bibliographystyle{apalike}
\bibliography{main}

\end{document}